\documentclass[manuscript]{aastex62}
\usepackage{natbib}

\usepackage{graphicx}
\usepackage[space]{grffile}
\usepackage{latexsym}
\usepackage{amsfonts,amsmath,amssymb}
\usepackage{url}
\usepackage[utf8]{inputenc}
\usepackage{fancyref}

\usepackage{multirow}
\usepackage{hyperref}
\received{}
\revised{}
\accepted{}
\shorttitle{Collisional processes in AGNs}
\shortauthors{Dimitrijevi{\' c} et al.}
\begin{document}

\title{THE ROLE OF SOME COLLISIONAL PROCESSES IN AGNS: RATE
COEFFICIENTS NEEDED FOR MODELING}

\correspondingauthor{Milan Dimitrijevi{\'c}}
\email{mdimitijevic@aob.rs}

\author{M. S. Dimitrijevi{\'c}}
\affil{Astronomical Observatory, Volgina 7, 11060 Belgrade 38, Serbia}

\author{V. A. Sre{\'c}kovi{\'c}}
\affiliation{Institute of Physics Belgrade, University of Belgrade, Pregrevica 118, 11080 Belgrade, Serbia}
%\nocollaboration

\author{Lj. M. Ignjatovi{\'c}}
\affiliation{Institute of Physics Belgrade, University of Belgrade, Pregrevica 118, 11080 Belgrade, Serbia}
%\nocollaboration

\author{B. P. Marinkovi{\'c}}
\affiliation{Institute of Physics Belgrade, University of Belgrade, Pregrevica 118, 11080 Belgrade, Serbia}



%\usepackage[T1]{fontenc}
%\usepackage{ae,aecompl}
%\usepackage{pdflscape}
%\usepackage{rotating,tabularx}
%\usepackage{lipsum}

%%%%% AUTHORS - PLACE YOUR OWN PACKAGES HERE %%%%%

% Only include extra packages if you really need them. Common packages are:
%\usepackage{graphicx}   % Including figure files
%\usepackage{amsmath}    % Advanced maths commands
%\usepackage{amssymb}    % Extra maths symbols
%\usepackage{lineno}
%\usepackage{longtable}
%\usepackage{rotating}
%\usepackage{color,soul}
%\usepackage{cases}
%\usepackage{booktabs}
%\usepackage{multirow}
%\usepackage{hyperref}

\begin{abstract}
The importance of some atom hydrogen collisions in AGN has
been investigated. The results are useful for better estimate of the
hydrogen Balmer lines fluxes, which usage for effective temperature
diagnostics in astrophysical plasma is limited by errors from the line
formation models. The data could be also useful for modeling cooler
and denser parts of AGN BLR clouds, as well as for the investigation of
Rydberg states of hydrogen and for the study of their influence during
the cosmological recombination epoch. The results of the present work
suggest that the investigated processes are of interest for the
research and modelling of such media.
\end{abstract}

\keywords{atomic data -- molecular data -- molecular processes-- atomic processes -- stars: atmospheres -- galaxies: active -- galaxies: nuclei}

%% \resthead is the RUNNING TITLE at top of the pages

\section{Introduction}
\label{sec:intro}

It has been shown in the paper of \citet{mih05}, on the basis of comparison with the relevant  electron-atom collisional processes,
that excitation/de-excitation processes in $\textrm{H}^*(n)+\textrm{H}(1s)$ collisions,
%%%%%%%%%%%%%%%%%%%%%%%%%%%%%%%%%%%%%%%%%%%%%%%%%%%%%%%%%%%%%%%%%%%%%%%%%%%%%%%
%\begin{equation}
%\textrm{H}^{*}(n) + \textrm{H} \rightarrow \left\{
%             \begin{array}{l}
%             \displaystyle{\textrm{H}^{*}(n'=n+p) + \textrm{H} },\\
%             \displaystyle{ \textrm{H} + \textrm{H}^{*}(n'=n+p)},
%             \end{array}
%     \right., \qquad n\ge 4, \quad p \ge 1,
%\label{eq:tri}
%\end{equation}
%%%%%%%%%%%%%%%%%%%%%%%%%%%%%%%%%%%%%%%%%%%%%%%%%%%%%%%%%%%%%%%%%%%%%%%%%%%%%%
%%and the inverse process of de-excitation
%%%%%%%%%%%%%%%%%%%%%%%%%%%%%%%%%%%%%%%%%%%%%%%%%%%%%%%%%%%%%%%%%%%%%%%%%%%%%%
%\begin{equation}
%\textrm{H}^{*}(n) + \textrm{H} \rightarrow \left\{
%             \begin{array}{l}
%             \displaystyle{\textrm{H}^{*}(n'=n-p) + \textrm{H} },\\
%             \displaystyle{ \textrm{H} + \textrm{H}^{*}(n'=n-p)},
%             \end{array}
%     \right. \qquad n-p \ge 4,
%\label{eq:cetiri}
%\end{equation}
%%%%%%%%%%%%%%%%%%%%%%%%%%%%%%%%%%%%%%%%%%%%%%%%%%%%%%%%%%%%%%%%%%%%%%%%%%%%%%
in the case of the principal quantum number $n \ge 4$, have important influence on the populations of hydrogen Rydberg atoms in weakly ionized parts
of the Solar photosphere and the
lower chromosphere. It was concluded that $(n-n')$-mixing processes should be taken into account in the modelling and research of these layers,
particularly around the temperature minimum in the solar photosphere.

In active galactic nuclei (AGN), especially in the region of the moderately ionized layers of dense parts of the broad-line region (BLR) clouds \citep{neg12,ili17,bon18} ($NeT \sim10^{14} \textrm{cm}^{-3}\textrm{K}$) plasma conditions are closer
to stellar atmospheres than to photoionized nebulae \citep{ost06,mar11,mar15}. We note that in spite of the fact that
irradiated part of the BLR clouds is highly ionized, they are enough big that temperature may decrease sufficiently
(for example up to around 2000 K, see Fig. 2 in \citet{cro93}).
%, where gas is weakly ionized and contains  a big number of hydrogen molecules \citep{cro93}.
Consequently, it is of interest to investigate the influence of the mentioned processes
in the cooler and denser parts of BLR clouds and to provide the data on the corresponding rate coefficients useful for modelling \citep{fer17} and investigations of such layers.
The usefulness of Balmer lines as an effective temperature diagnostics is potentially limited by errors in the line formation models and uncertainties in used atomic/molecular data for hydrogen and inelastic collisions \citep{ili12, ama18},
so that this topic is important and current.

Collisional processes are not so significant in photoionized galactic nebulae, since there, the density is low and
most of atoms, ions and molecules are in the ground state. The situation is different in some parts of the broad line clouds where many atoms
are in excited states,
because of the high density and large optical depth \citep{neg12,net13}. In the BLR clouds, the excited states $n \gg 1$ of hydrogen, helium and some metals and their populations are completely controlled by such processes
\citep[see e.g.][]{bla90,net13}. The theoretical calculations for these processes are not of great accuracy, since a very large number of different influences should be taken into
account, and many unknown cross sections must be guessed as well as
uncertainties in the known ones \citep{bar07}. Some of them must be updated in atomic/molecular databases \citep{dub16,lav18a} and checked before they will be used as input parameters in simulation codes (see paper of \citealt{sre18a} with
new data of $\textrm{H}^*+\textrm{H}$ collisions and e.g. new data for $\textrm{H}^-+\textrm{H}$ in \citealt{gro19}, etc.). This is sufficient reason for further investigation of these possibly important collisional process in the very dense parts of the BLR (see e.g. \citealt{fer17,lav18b}).
%\subsection{Collisional excitation}

In recent paper of \citet{sre18a} it was demonstrated that the ionization processes in $\textrm{H}^*(n)+\textrm{H}(1s)$ collisions and also the inverse recombination processes, influence on the ionization level and atom excited-state populations in weakly ionized regions in the hydrogen clouds in broad-line region of active galactic nuclei (AGNs), and must have influence on the optical properties.
Here we want to examine further and to point out the importance of the $(n-n')$-mixing processes in some dense moderately ionized parts in broad-line region (BLR) of AGNs. That's why we have extended the investigation of these processes and calculated their rate coefficients for a wider region of plasma parameters and for quantum numbers $n$ up to 20.
That highly excited Rydberg atoms are important because e.g. radio recombination lines (RRL) coming from highly excited Rydberg levels can be used to determine the densities and temperatures of gaseous nebulae (see e.g. \citet{ost06}). Peculiarly, hydrogen RRL can be used to obtain the electron temperature.

\section{Collisional excitation/deexcitation processes}
This paper is the next step of our investigations of $\textrm{H}^*(n)+\textrm{H}(1s)$  collision processes, that
influence the excited hydrogen populations in moderately ionized layers. In principle, there are two channels of these processes.
The first is the chemi-ionization processes \citep{mih11,mih12,sre18b}, connecting the block of atomic Rydberg states with the continuum. The second is the $(n-n')$-mixing processes i.e. the excitation and deexcitation processes \citep{mih04,mih08,sre13}, that indicate transitions between the Rydberg states with the principal quantum numbers $n$ and $n' \ne n$.

Mihajlov and coworkers demonstrated that the Rydberg state distribution in a weakly ionized hydrogen plasma may be as well strongly
influenced by $(n-n')$-mixing processes in $\textrm{H}^{*}(n) + \textrm{H}$ collisions i.e. excitation processes
%%%%%%%%%%%%%%%%%%%%%%%%%%%%%%%%%%%%%%%%%%%%%%%%%%%%%%%%%%%%%%%%%%%%%%%%%%%%%%
\begin{equation}
\textrm{H}^{*}(n) + \textrm{H} \rightarrow \left\{
             \begin{array}{l}
             \displaystyle{\textrm{H}^{*}(n'=n+p) + \textrm{H} },\\
             \displaystyle{ \textrm{H} + \textrm{H}^{*}(n'=n+p)},
             \end{array}
     \right., \qquad n\ge 4, \quad p \ge 1,
\label{eq:tri}
\end{equation}
%%%%%%%%%%%%%%%%%%%%%%%%%%%%%%%%%%%%%%%%%%%%%%%%%%%%%%%%%%%%%%%%%%%%%%%%%%%%%
and the inverse process of de-excitation
%%%%%%%%%%%%%%%%%%%%%%%%%%%%%%%%%%%%%%%%%%%%%%%%%%%%%%%%%%%%%%%%%%%%%%%%%%%%%
\begin{equation}
\textrm{H}^{*}(n) + \textrm{H} \rightarrow \left\{
             \begin{array}{l}
             \displaystyle{\textrm{H}^{*}(n'=n-p) + \textrm{H} },\\
             \displaystyle{ \textrm{H} + \textrm{H}^{*}(n'=n-p)},
             \end{array}
     \right. \qquad n-p \ge 4,
\label{eq:cetiri}
\end{equation}
%%%%%%%%%%%%%%%%%%%%%%%%%%%%%%%%%%%%%%%%%%%%%%%%%%%%%%%%%%%%%%%%%%%%%%%%%%%%%
provoqued by the same resonant mechanism as the processes of chemi-ionization and chemi-recombination. These results
demonstrate the need of an investigation of the processes (\ref{eq:tri}) and (\ref{eq:cetiri}) in weakly ionized regions of
stellar atmospheres \citep{mih05,mih05b}. The influence of the $(n-n')$-mixing processes have also been analyzed in \citet{bar07} and \citet{mas09}.
The reason for further investigation of these processes is the existing uncertainties in the rate coefficients due to hydrogen collisions in many cases as concluded in \citet{bar07}. Recently in \citet{guz19} authors show that at intermediate to high densities, electron excitation/deexcitation collisions are the dominant process for populating or depopulating high Rydberg states and the accurate knowledge of the energy  collisional rates is determinant for predicting the radio recombination spectra of gaseous nebula.

For the mentioned reasons we shell investigate in more detail the importance of
excitation/de-excitation processes in $\textrm{H}^*(n)+\textrm{H}(1s)$ collisions
(\ref{eq:tri}) and (\ref{eq:cetiri}) in the moderately ionized layers of cooler and denser parts of the BLR clouds, with plasma conditions closer to stellar
atmospheres than photoionized nebulae \citep{ost06}. Consequently, we shall calculate the rate coefficients of these processes for
different $n$ and $p$ within the frame of the method developed and presented in \citet{mih04,sre13}. The obtained rate coefficients will be compared with the rate coefficients of some concurrent processes e.g. electron-atom mixing
%%%%%%%%%%%%%%%%%%%%%%%%%%%%%%%%%%%%%%%%%%%
\begin{equation}\label{eq:electron_impact_mixing}
e+\textrm{H}^{*}(n)\rightarrow e+\textrm{H}^{*}(n').
\end{equation}
%%%%%%%%%%%%%%%%%%%%%%%%%%%%%%%%%%%%%%%%%%%

%%%%%%%%%%%%%%%%%%%%%%%%%%%%%%%%%%%%%%%%%%%%%%%%%%%%%%%%%%%%
\begin{figure*}
\includegraphics[width=1\columnwidth]{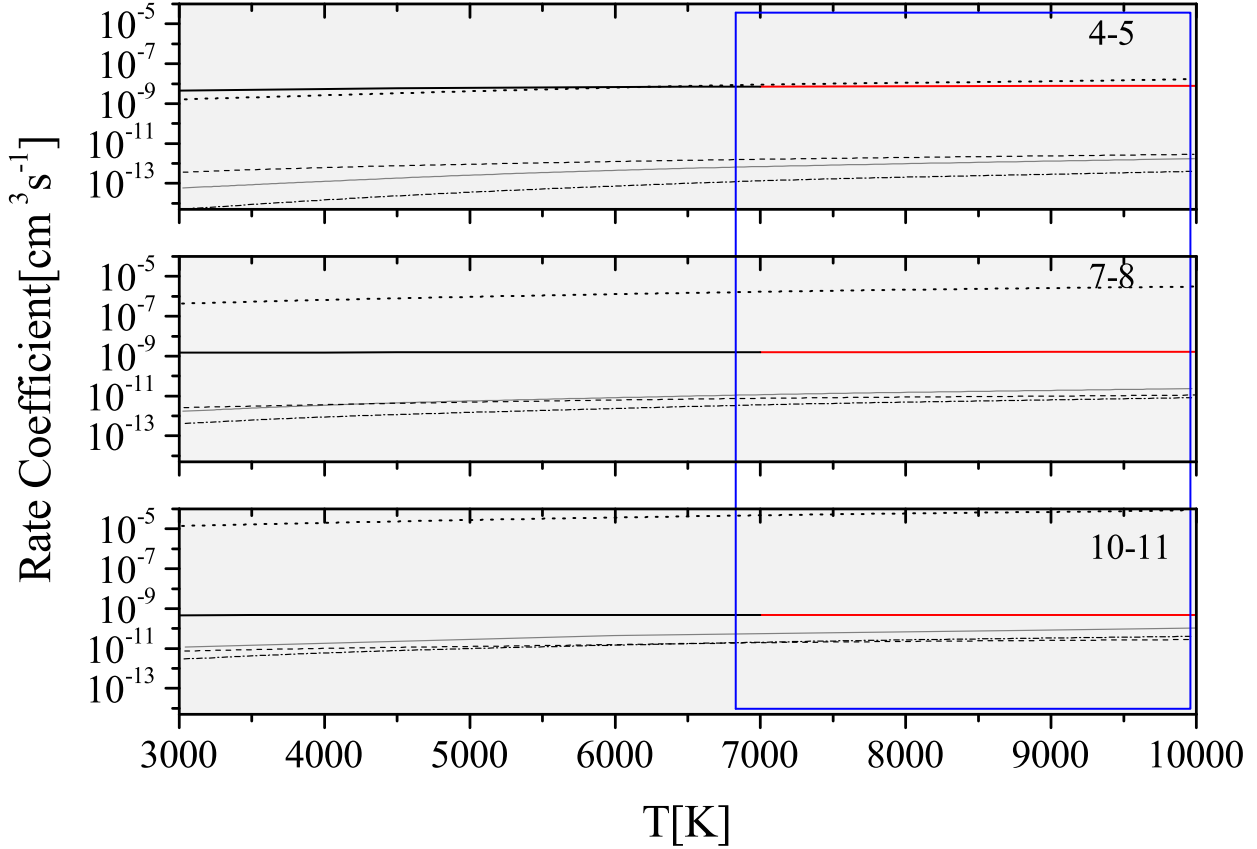}
\caption{\label{fig:bark} Plot of excitation rate coefficients for selected excited states and temperatures important for physics of AGN BLR clouds.
The black lines are the data analyzed in \citet{bar07}. The data from Mihajlov and coworkers based on the same mechanism as here
are plotted as normal full lines. The red lines are excitation rate coefficients calculated in this work for $T \ge 7000K$ (blue marked region).
The numerical  data from \citet{soo92} are plotted as thick full lines and the dot-dashed
line, the analytic data from Soon are plotted as dashed lines, and the data from \citet{dra68,dra69} are plotted as dotted lines. }
\end{figure*}
%%%%%%%%%%%%%%%%%%%%%%%%%%%%%%%%%%%%%%%%%%%%%%%%%%%%%%%%%%%%
For a collision of an electron with atomic hydrogen which is represented by Eq.\ref{eq:electron_impact_mixing} where symbols $n$ and $n'$ stand for sets of quantum numbers $(n, l, m)$ and $(n’, l’, m’)$, respectively, there are several processes that can be considered as elastic scattering, excitation and de-excitation (including spin-flip), ionization, and radiative electron capture. All these processes are of great importance for astrophysics, especially studies in the stellar physics, determination of spectral line profiles \citep{dim14}, or in the physics of the interstellar gas.

Beside differential and integral cross sections for each of abovementioned processes, many applications need the collisional rate coefficients that are the temperature-averaged cross sections integrated over the complete energy dependence.
\citet{vri80} gave a number of practical formulas, applicable in a wide range of incident electron energies, which included transitions either from the ground state or from highly excited states. Considering Rydberg states the usual assumption
is that their distribution is determined by electron collisions \citep{mih05}. \citet{agg91} published collision strength datasets for all hydrogen transitions $n \rightarrow n’$ between the levels from 1 to 5 at energies below 26.7 eV by the
use of R-matrix and no-exchange calculations. The calculations by \citet{bra92} who used the converged close coupling (CCC) method and by \citet{bar96} who used the R-matrix with pseudo-states (RMPS) method to calculate the three lowest transitions for incident electron energies between 10.23 eV and 12.08 eV i.e. the $n =$ 2 and 3 thresholds are considered as the benchmark data for electron–hydrogen scattering.  \citet{bar06} performed an extensive computational study of electron-impact scattering and ionization of atomic hydrogen and hydrogenic ions, in non-relativistic frame by using the propagating exterior complex scaling (PECS) method applying it at the broad range of incident energies. Recently, \citet{ben18} published converged datasets containing scattering data for collisions of electrons on the atomic hydrogen including magnetic sub-levels for total energies below the $n = 4$ excitation threshold. They used the direct solution of the Schr\"{o}dinger equation in the B-spline basis with the exterior complex scaling (ECS) boundary condition. For astrophysical processes the low energies (low temperatures) are of great importance and that is why these simulations are rigorously done right above the excitation thresholds, with the resolved all major resonances.
The data sets are available through an URL link
\href{url}{http://utf.mff.cuni.cz/data/hex}. Some electron differential cross section data for elastic scattering and electron excitation by hydrogen are also provided by the Belgrade Electron/Atom and Molecule Database (BEAMDB) \citep{mar17, mar19}.

The question of hydrogen molecule formation within quasar BLR clouds has been considered by developing models for a numerous both radiative and collisional processes with basic constituents \citep{cro93} using the available cross sections/rate
coefficients, while the efficient model that considers the presence of physisorbed and chemisorbed sites on the dusty grains with allowing surface diffusion by quantum mechanical tunneling and thermal hopping for absorbed H atoms was developed
by \citet{caz02}. Looking at the collisional processes we can divide those in two classes, either lepton (electron) or hadron (proton, neutral or ion) collisions. The most studied are the processes of electron scattering by hydrogen molecule that
include elastic scattering, vibrational and electronic excitations, ionisation, dissociative electron attachment or direct dipolar dissociation. It is interesting to note that for the processes like the associative
ionization, $\textrm{H}(ls) + \textrm{H}(2s) \rightarrow \textrm{H}_2^+ + e^-$ \citep{urb91}, ion pair production in $\textrm{H}_A(2s) + \textrm{H}_B(ls)$ collisions leading to direct $(\rightarrow \textrm{H}_A^+ + \textrm{H}_B^-)$ and
indirect $(\rightarrow \textrm{H}_B^+ + \textrm{H}_A^-)$ charge exchange \citep{fus82}, or low energy associative detachment, $\textrm{H}^-(^1S) + \textrm{H}(^2S) \rightarrow \textrm{H}_2(v,j) + e^-$ \citep{gro19} exist inverse processes
and their cross sections could be studied or compared with such inverse cross sections for electron collisions. In the database BEAMDB \citep{mari15} the electron cross sections for hydrogen molecule (dihydrogen) are represented by datasets
for elastic scattering \citep{sri75}, electronic excitation of the $a^1 \Sigma_g^+$, $B^1 \Sigma_u^+$, $c^3 \Pi_u$ and $C^1 \Pi_u$ states \citep{kha86} and the $b^3 \Sigma_u^+$ state \citep{kha87}.

The calculated rate coefficients of the corresponding processes are presented in tables and figures. The tables cover the regions
$4 \le n \le 20$ and $2000 \textrm{ K} \le T \le 30000 \textrm{ K}$, much wider than the regions relevant for moderately
ionized layers of dense parts of the BLR clouds \citep[see e.g.][]{neg12, cro93}, so that these data can be used and for other purposes like,
the atmospheres of Sun and Sun-like stars, for the investigation of Rydberg states of hydrogen and for the study of cosmological recombination epoch \citep{gne09,mih11,chl10}.

%  *********   Start Figure 2  ****************
\begin{figure}
\begin{center}
\includegraphics[width=0.7\columnwidth,
height=0.5\columnwidth]{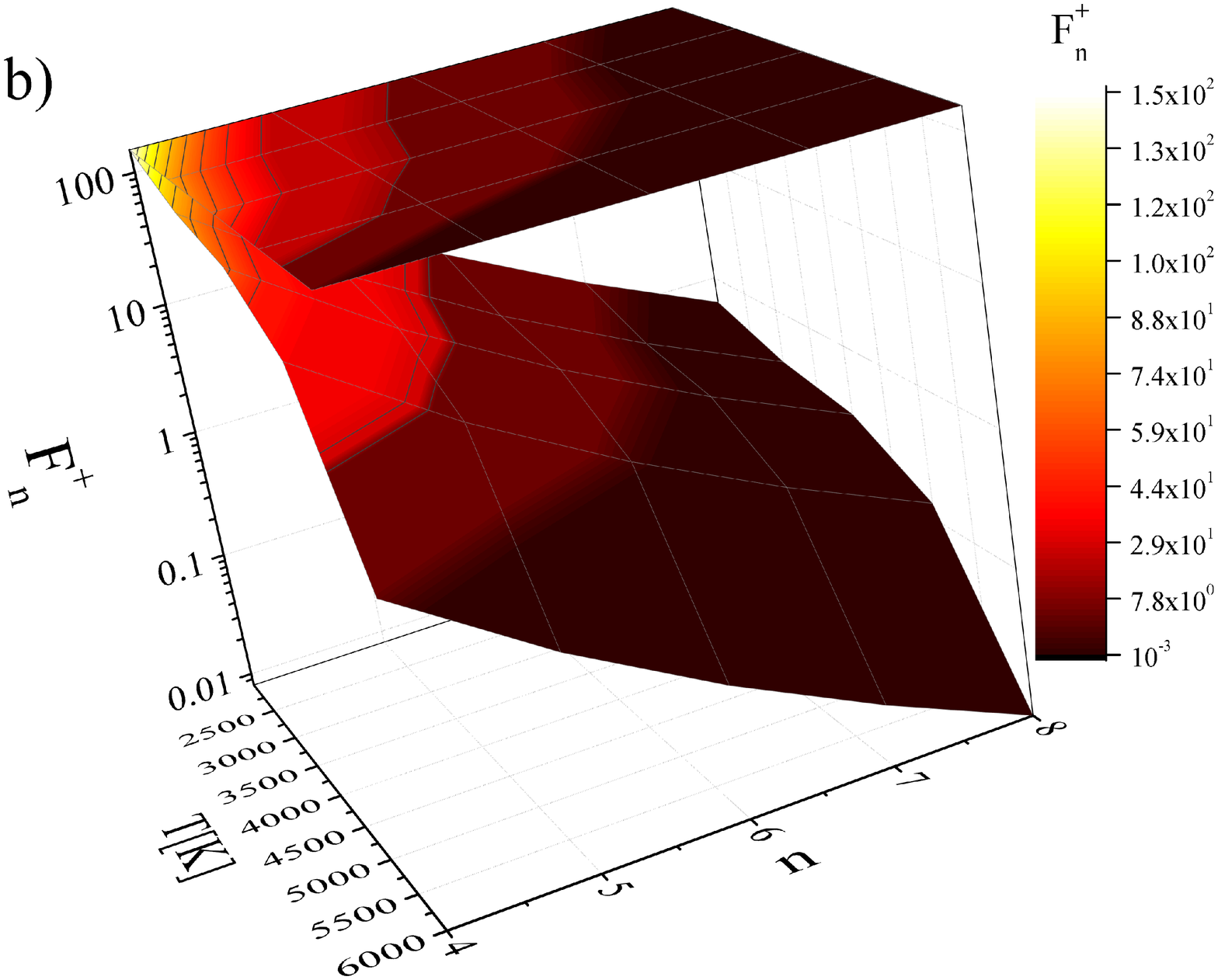}
\includegraphics[width=0.7\columnwidth,
height=0.5\columnwidth]{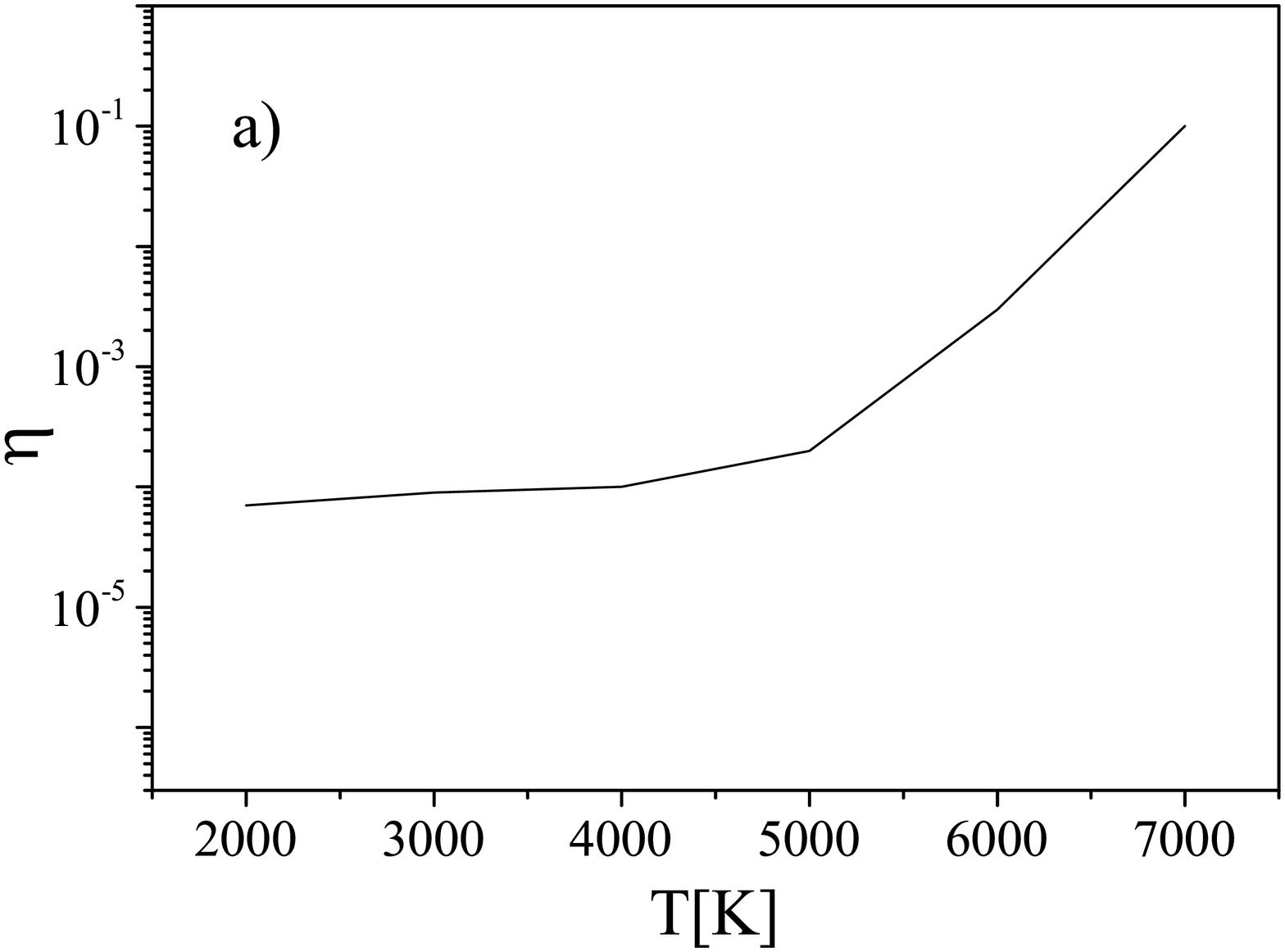}
\caption{a) : The coefficient $\eta$ for $2000 \textrm{ K} \le T \le 7000 \textrm{ K}$, obtained using Saha equation;
 b): A surface plot of the quantities Fn(+) defined by Eq.\eqref{eq:F+-} with $4\le n\le 8$, for $2000 \textrm{ K} \le T \le 6000 \textrm{ K}$.}
\label{fig:DB1}
\label{fig:DB1}
\end{center}
\end{figure}
%  *********   End Figure 2  ****************

\section{Resonant mechanism}
The resonant mechanism, used in this paper for processes (\ref{eq:tri}) and (\ref{eq:cetiri}) has been explained in a couple of previous papers and reviewed in details in  \cite{mih12}, so that only the basic description will be given here. The resonant mechanism is used for the $\textrm{H}^{*}(n) + \textrm{H}$ collision system  within the region $R \ll r_{n}$ where $R$ is the internuclear distance and $r_{n} \sim n^{2}$ is the average radius of the  $\textrm{H}^*(n)$ atom.  Within this region the $\textrm{H}^{*}(n)+\textrm{H}$ system is described as: $e + (\textrm{H}^{+}+\textrm{H})$, where $e$ is the outer electron of the $\textrm{H}^*(n)$ atom. In order to describe the electronic states of the subsystem  $(\textrm{H}^{+}+\textrm{H})$, the adiabatic electronic ground state, or the first excited state of the molecular ion $\textrm{H}_{2}^{+}$ were used.

The processes (\ref{eq:tri}) and (\ref{eq:cetiri}) of $(n-n')$-mixing, similar as the processes of chemi-ionization/recombination,
are described as a result of the resonant energy exchange between the outer electron $e$ and the electronic component of the
$\textrm{H}^{+}+\textrm{H}$ subsystem \citep{sre18a}.
The transition of the outer electron from the initial  to the upper energetic states occurs together with the transition
of the $\textrm{H}^{+}+\textrm{H}$ subsystem from the electronic excited state  to the ground state, while the transition of the
outer electron to the lower energetic state is simultaneous with the transition of the $\textrm{H}^{+}+\textrm{H}$ subsystem from
the ground electronic state to the excited state. The mentioned transitions occur due to the interaction of the outer electron
with the dipole momentum of the $\textrm{H}^{+}+\textrm{H}$ subsystem. The described resonant mechanism has been applied as in \citep{mih05} for the principal quantum number region $n \ge 4$.

The processes (\ref{eq:tri}) and (\ref{eq:cetiri}) are characterized by the excitation and deexcitation rate coefficients $K_{n;n+p}(T)$ and $K_{n;n-p}(T)$, where $T$ is the local temperature of the atomic particles, determined with a method similar to that used in \citet{sre13} or \cite{mih08}. Namely in the mentioned articles the excitation rate coefficients $K_{n;n+p}(T)$ have been calculated directly and numerically semi-classically, and the deexcitation rate coefficients $K_{n;n-p}(T)$ have then been determined with the help of the principle of thermodynamical balance. All needed relations may be found in  \citet{mih08}.

To estimate the relative efficiency of processes (\ref{eq:tri}) and (\ref{eq:cetiri}) in comparison to process (\ref{eq:electron_impact_mixing}) we
calculated the quantities
\begin{equation}
\label{eq:F+-}
F_{n\pm}= \frac{\sum_{p=1}^{5}K_{n;n \pm p}(T)N(n)N(1)}{\sum_{p=1}^{5}\alpha_{n;n \pm p}(T)N(n)N_{e}}=\frac{\sum_{p=1}^{5}K_{n;n \pm p}(T)}{\sum_{p=1}^{5}\alpha_{n;n \pm p}(T)}\cdot \eta.
\end{equation}
Here $N(n)$ is the excited atom states population for given $n$, $N_{e}$ is electron density, $\eta=(N(1)/N_{e})$, $\alpha_{n;n \pm p}(T_{e}=T)$ is the rate coefficient for the electron-atom process (\ref{eq:electron_impact_mixing}) taken from \citet{vri80} and \citet{joh72}. With $T_{e}$ is denoted the electron temperature, generally not equal to the atomic temperature. The products $K_{n;n \pm p}(T)N(n)N(1)$ and $\alpha_{n;n \pm p}(T)N(n)N_{e}$ are the partial atom- and electron-Rydberg atom excitation/deexcitation fluxes.
%To estimate the relative efficiency of processes (\ref{eq:tri}) and (\ref{eq:cetiri}) in comparison to process (\ref{eq:electron_impact_mixing}) we
%calculate ratio of corresponding fluxes $(\sum_{p=1}^{5}K_{n;n \pm p}(T)N(n)N(1))/(\sum_{p=1}^{5}\alpha_{n;n \pm p}(T)N(n)N_{e})=$ $(\sum_{p=1}^{5}K_{n;n \pm p}(T))/(\sum_{p=1}^{5}\alpha_{n;n \pm p}(T))\cdot(N(1)/N_{e})$.
%Here $N(n)$ is the excited atom states population for given $n$, $\alpha_{n;n \pm p}(T_{e}=T)$ is the rate coefficient for the electron-atom process (\ref{eq:electron_impact_mixing}) taken from \citet{vri80} and \citet{joh72}. With $T_{e}$ is denoted the electron temperature, generally not equal to the atomic temperature. The products $K_{n;n \pm p}(T)N(n)N(1)$ and $\alpha_{n;n \pm p}(T)N(n)N_{e}$ are the partial atom- and electron-Rydberg atom excitation/deexcitation fluxes.
We should add that ion-atom non elastic collisional processes \citep{mih13, sre14}, in spite of the fact that they are also characterized by long-range interaction, here are not of significance. This is due to the very large  difference of masses of electron and ion, so that the impact ion-atom velocity is several orders of magnitudes lower than electron-atom impact velocity.

\section{Results and Discussion}
\label{sec:results}

The rate coefficients $K_{n;n \pm p}(T)$ for processes (\ref{eq:tri}) and (\ref{eq:cetiri}) are calculated in the
domains of $n$ and $T$ corresponding to the conditions in moderately ionized hydrogen plasmas, of interest for dense parts of BLR clouds in AGNs. The values for excitation rate coefficients $K_{n;n + p}(T)$ with $4 \le n \le 20$, $1 \le p \le 5$
and $2000\textrm{ K} \le T \le 30000\textrm{ K}$ are presented
in the tables (Tabs. 3–20, Supplementary material) in the online version of this article. The sample of the results is provided in Tab.\ref{tab:table1} in order to demonstrate
the content of additional data and their form.
%%%%%%%%%%%%%%%%%%%%%%%%%%%%%%%%%%%%%%%%%%%%%%%%%%%%%%%%%%%%%%%%%%%%%%%%%%%
\begin{table}
\begin{center}
\caption{Excitation rate coefficients $K_{n;n + p}(T)$ ($10^{-9} \textrm{cm}^{3}\textrm{s}^{-1}$). A portion is shown here for guidance regarding its form and content.} \label{tab:table1}
\begin{tabular}{ccccccc}
\\ \hline
%This is the header for the remaining page(s) of the table...
\hline \\
\multicolumn{7}{c}{T[K]}\\
                    \cline{3-7}
    n    & p & 2000 & 8000   &  10000   & 20000  & 30000  \\
\hline
\multirow{5}{*}{12} & 1 & 0.24269 & 0.24715  & 0.24738 &  0.24781  & 0.24794 \\
                    & 2 & 0.08782 & 0.09212  & 0.09236 &  0.09281  & 0.09295 \\
                    & 3 & 0.04653 & 0.05005  & 0.05025 &  0.05063  & 0.05075 \\
                    & 4 & 0.02881 & 0.03165  & 0.03182 &  0.03213  & 0.03223 \\
                    & 5 & 0.01951 & 0.02183  & 0.02196 &  0.02222  & 0.02230 \\
                    \hline
\end{tabular}
\end{center}
\end{table}
%%%%%%%%%%%%%%%%%%%%%%%%%%%%%%%%%%%%%%%%%%%%%%%%%%%%%%%%%%%%%%%%%%%%%%%%%%%%%%%%%%%%%%%%%%%

These results for temperatures $\le 10000 $K are relevant for the conditions in moderately ionized layers of dense parts of the BLR clouds
\citep[see][]{neg12, cro93}, and enable the potential inclusion of these processes in their modeling and investigations.
The reason that we present and data for higher temperatures is since they may be of interest for other moderately ionized plasmas, as well as for the investigation of Rydberg states of hydrogen and for the study of their influence during the cosmological
recombination epoch \citep[see][]{chl10}. The Rydberg states that arise have large dipole moments, leading to strong absorption in the infrared part of spectra,
and the appearance of polarization \citep{gne09,afa18}.
%Radio recombination lines (RRL) coming from highly excited Rydberg levels can be used to determine the densities and temperatures of gaseous nebulae (see e.g. \citet{ost06}).

We give the rate coefficients as a second degree polynomial fit to numerical results over the temperature range of $2000 \textrm{ K} \le T \le 30 000$ K (Tab. 20 online). In Tab.~ \ref{tab:fit} are presented the selected fits (for $n = 4,6,8,10$ and $p=1$ i.e. n'=n+1).

In Fig. \ref{fig:bark} the comparison of excitation rate coefficients for selected excited states is presented for the conditions important for physics of AGN BLR clouds. The black lines are the data from the existing literature analyzed in \citet{bar07}, in the narrow parameter region ($T\le 10000 \textrm{ K}$
and $n \le 10$). The data from Mihajlov and coworkers based on the same mechanism as in this work are plotted as full lines. The red lines are rate coefficients obtained in this work. One can see the present uncertainties on the rate coefficients due
to hydrogen collisions in many cases as concluded in \citet{bar07}.

\begin{figure}
\includegraphics[width=0.8\columnwidth,
height=0.6\columnwidth]{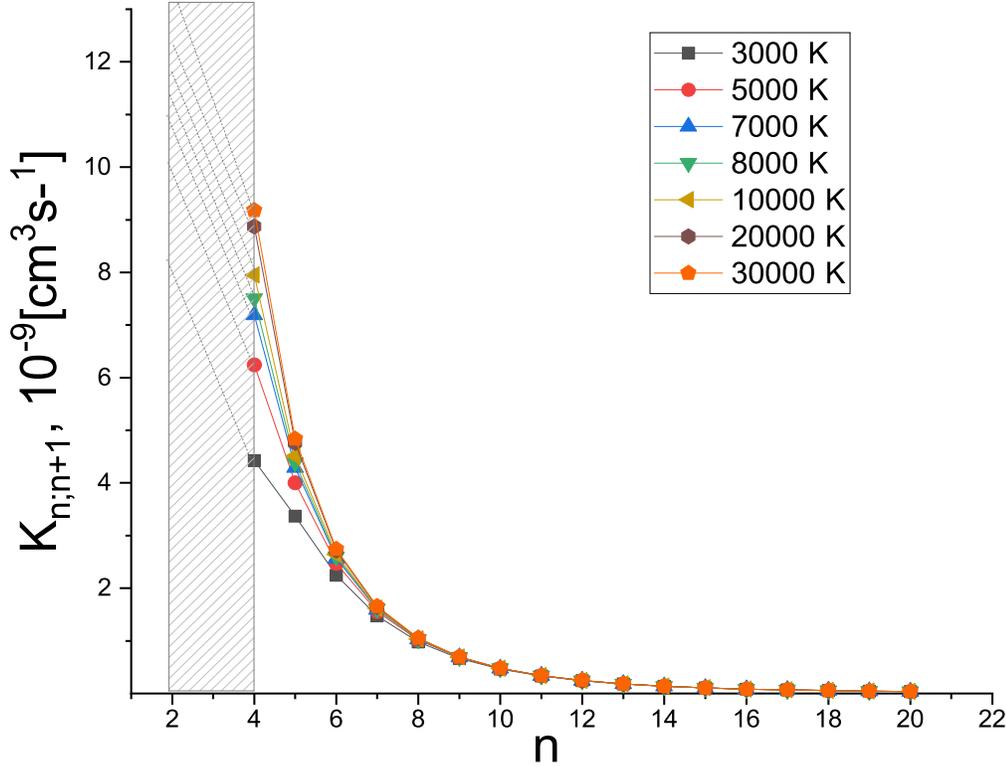}
\caption{\label{fig:K} Plot of excitation rate coefficients $K_{n;n + 1}(T)$ for excited states
$4 \le n \le 20$, $p=1$ i.e. $n'=n+1$ and $T \le 30000\textrm{ K}$. In the left shaded
region the extrapolated values are presented.}
\end{figure}

Although the determination of rate coefficients for parameters that exists in BLR clouds was our primary task in order to be used in the corresponding models we have, also examined the impact of these processes. We compared the relative influence of $(n-n')$-mixing processes (\ref{eq:tri}) and (\ref{eq:cetiri}) and  influence of concurrent electron-atom excitation processes (\ref{eq:electron_impact_mixing}) on the same block of excited hydrogen atom states with $4\le n \le 10$. Fig. \ref{fig:DB1} illustrates the behavior of the ratio of corresponding fluxes for the conditions that exist  in BLR of AGNs. Using the Saha equation we obtained the density ratios of electrons and hydrogen atoms as shown in Fig. \ref{fig:DB1}a. In the lower temperature region it has been confirmed the domination of the $(n-n')$-mixing processes for $ n \le 6$ over the mentioned concurrent processes while for $7 \le n\le 10$ they are comparable with the concurrent excitation processes (see Fig \ref{fig:DB1}b). For the parts of BLR with higher temperatures  processes (\ref{eq:tri}) and (\ref{eq:cetiri}) are comparable with concurrent excitation processes (\ref{eq:electron_impact_mixing}) only for lower $n$.
For the area $2 \le n < 4$ reliable data for the corresponding rate coefficients are missing.
The rate coefficients for $n=2$ and 3 are data from \cite{dra68}, which only exist in literature, but they are not usable (see papers of \citet{bar07,mas09}). The fact and the big problem is that this area is of great importance and it raises many issues such as the conclusion that their impact should be much much higher (see shaded part of Fig.~\ref{fig:K}). Also, important fact is that the uncertainties  on  the  excitation  rates  at  high  Rydberg  levels could certainly affect the lower levels by cascading decay, and thus compete with the errors from the collisional data for low-lying levels \citep{guz19}.
%%%%%%%%%%%%%%%%%%%%%%%%%%%%%%%%%%%%%%%%%%%%%%%%%%%%%%%%%%%%%%%%%%%%%%%%%%%%%
\begin{table}
	\centering
	\caption{The fits $\log (K_{n;n+p}(T))= \sum_{i=0}^{2} k_{i} (\log (T))^{i}$ to the rate coefficient. A portion is shown here for guidance regarding its form.}
	\label{tab:fit}
\begin{tabular}{cccc} \hline \hline
$n-n'$  & $k_0$       & $k_1$       & $k_2$         \\
\hline
4-5   & -15.1815 & 3.259150 & -0.3719690  \\
6-7   & -10.5293 & 0.899118 & -0.1028000    \\
8-9   & -9.69682 & 0.328757 & -0.0376158 \\
10-11 & -9.61839 & 0.137414 & -0.0157439 \\
\hline
\end{tabular}
\end{table}
%%%%%%%%%%%%%%%%%%%%%%%%%%%%%%%%%%%%%%%%%%%%%%%%%%%%%%%%%%%%%%%%%%%%%%%

The data demonstrate the fact that the considered $(n-n')$-mixing processes (\ref{eq:tri}) and (\ref{eq:cetiri}) must have a noticeable influence on the populations of excited hydrogen atoms in cooler and denser parts of the BLR clouds in AGNs in
comparison to the concurrent processes. We note that one reason for further investigation of these processes is the present uncertainties of the rate coefficients due to hydrogen collisions especially for higher $n$ (two order of magnitude differences in rate coefficients) as concluded in \citet{bar07} and we assume that their impact should be higher.
It is obvious that the importance of these processes for modeling of moderately ionized layers of dense parts of the BLR clouds should be necessarily investigated and that they should be included in the databases and standard models \citep{lav18a,fer17}. Apart from that, the obtained results could be also useful for modeling of different stellar atmospheres \citep{prz04,bar07,mas09,fon09},
as well as for the investigation of Rydberg states of hydrogen \citep{gne09, guz19}, for the study of their influence during the cosmological recombination epoch \citep[see e.g.][]{chl10} and for simulation of the formation of massive seed black holes in the early Universe \citep{glo15}.
%%%%%%%%%%%%%%%%%%%%%%%%%%%%%%%%%%%%%%%%%%%%%%%%%%%%%%%%%%%%
%%%%%%%%%%%%%%%%%%%%%%%%%%%%%%%%%%%%%%%%%%%%%%%%%%%%%%%%%%%%

\section{Conclusions}
\label{sec:conclusions}
We calculated (n-n') mixing rate coefficients $K_{n;n + p}(T)$ in $\textrm{H}^*(n)+\textrm{H}(1s)$ collisions for $4 \le n \le 20$, $1 \le p \le 5$
and $2000\textrm{ K} \le T \le 30000\textrm{ K}$ (Tabs. 3–20). The  rate coefficients data of the corresponding processes
are given in tabulated form easy for further use in modeling of analysed environments.
We estimated the relative efficiency of (n-n') mixing processes in comparison to electron-atom mixing process by calculating the quantities Fn. From these considerations we found that in the region of temperatures between 2000 K and 6000 K the density ratios of electrons vs hydrogen atom is just $\sim 10^{-2} - 10^{-3}$ and might be of relevance for the investigation of cooler and denser parts of BLR clouds. It will be very useful to perform an analysis for example with the
code CLOUDY \citet{fer17} in order to see changes in optical characteristics.
%using  the  spectral simulation code CLOUDY \citep{fer17}.

%\vspace{-6pt}
\
\

This work has been supported by the Ministry of Education, Science and Technological Development of the Republic of
Serbia Grants OI176002, III44002, OI171020.

\section{SUPPORTING INFORMATION}
Additional Supporting Information (Tabs. 3 - 20) may be found in the on-line version of this article.
The tables are available in its entirety for $4 \le n \le 20$, and $2000 \textrm{ K} \le T \le 30000 \textrm{ K}$
in machine-readable form in the online journal as additional data.

\end{document}